# An Adaptable System to Support Provenance Management for the Public Policy-Making Process in Smart Cities

**Barkha Javed \*, Zaheer Khan and Richard McClatchey**

Department of Computer Science and Creative Technologies, University of the West of England, Bristol BS16 1QY, UK; Zaheer2.Khan@uwe.ac.uk (Z.K.); Richard.Mcclatchey@uwe.ac.uk (R.M.)

**\*** Correspondence: barkha.javed@uwe.ac.uk

**Abstract:** Government policies aim to address public issues and problems and therefore play a pivotal role in people's lives. The creation of public policies, however, is complex given the perspective of large and diverse stakeholders' involvement, considerable human participation, lengthy processes, complex task specification and the non-deterministic nature of the process. The inherent complexities of the policy process impart challenges for designing a computing system that assists in supporting and automating the business process pertaining to policy setup, which also raises concerns for setting up a tracking service in the policy-making environment. A tracking service informs how decisions have been taken during policy creation and can provide useful and intrinsic information regarding the policy process. At present, there exists no computing system that assists in tracking the complete process that has been employed for policy creation. To design such a system, it is important to consider the policy environment challenges; for this a novel network and goal based approach has been framed and is covered in detail in this paper. Furthermore, smart governance objectives that include stakeholders' participation and citizens' involvement have been considered. Thus, the proposed approach has been devised by considering smart governance principles and the knowledge environment of policy making where tasks are largely dependent on policy makers' decisions and on individual policy objectives. Our approach reckons the human dimension for deciding and defining autonomous process activities at run time. Furthermore, with the network-based approach, so-called provenance data tracking is employed which enables the capture of policy process.

## 1. Introduction

The term "smart cities" has attracted wide attention and is defined differently by various researchers [1–3], yet to date no single widely accepted definition of a smart city exists. Having no agreed definition of smart cities often leaves its interpretation to the body responsible for controlling and managing cities' operations. However, this ambiguity also gives cities the liberty to select the smart city project that targets their specific objectives, concerns and problems [4]. Our research investigates government operations pertaining to the policy creation process; consequently, this research focuses on the "smart government" dimension of smart cities. Similar to smart cities, there is no one agreed definition of smart government [5–9] and it often encompasses governments employing ICT for improving citizens' lives [5], involving various stakeholders for the management and implementation of policies [6], the consideration of smart governance practices by government [7], the improvement of city management and city services [8], or the use of ICT to gather city data in order to improve the quality of citizens' lives [9].

Different initiatives, such as transforming e-government to smart government [10], have been carried out to provide smarter government, however, what has not been considered is a technology that assists government during policy formulation. Although a philosophical stance on the suitability of workflow technology has been put forward by Sajjad [11], its practical implication has not been investigated. The research presented in this paper therefore focuses on proposing a technological solution that assists policy makers (bodies which are involved in public policy decision making) in their operations associated with policy formation such as communication and collaboration, routing tasks, facilitating citizens' participation, information generation as part of the process and the storing and organization of data [11,12]. Further study [13] signifies the potential of previously collected policy data for improving the implementation of further policies; for example, it shows that the analysis of budgetary patterns of previous policies can provide useful feedback regarding budgets for new policies. Additionally, Van der Aalst [14] shows that data, produced as a result of business processes execution, can serve as an important asset for improving and re-engineering the business process. What is missing at present is a tracking service (scientifically, this tracking is often termed the "provenance" or history of the process and its establishment) that enables policy process tracking. This provenance capture in the policy making process requires ICT support; consequently, our research focuses on proposing a solution to assist in the execution of policy and in tracking the process data that acts as the foundation for policy creation. The aim of this work is to design a system that can facilitate the provenance management of the policy creation process. Here provenance management refers to provenance data capture, its storage and its retrieval. However, the scope of this paper is limited to operations related to the provenance data capture mechanisms i.e. how ICT facilitates provenance capturing for policy making. The full details of provenance, its retrieval and usage are outside the scope of this paper.

From a technical perspective, suitable technologies are needed to capture the provenance in a policy making process, for example workflow technology [11]. A detailed in-depth exploration of the policy process has revealed (details are presented in Section 2) that tracking policies is not without challenge; this is mainly due to the inherent nature of policy processes [11]. What is required is a process-agnostic approach that does not simply rely upon process design and orchestration (e.g. workflows) but enables "just-in-time" enactment of policy creation activities. This is because having a pre-defined process for setting up policy is complex and may not be suitable since it requires: (i) frequent modifications in processes in order to address the requirements of various policies (ii) the identification of process activities, which can be challenging for policy setup where the next activity execution largely depends on human knowledge and experience (iii) the assignment of particular activities to roles which also imparts an overhead of identifying well-defined process activities and their sequence; a large number of stakeholders further exacerbates this challenge and (iv) guiding human decision and activities with a well-defined process is practically non-ideal for a knowledge intensive environment of a policy setup.

Considering these challenges, a novel idea based on the familiar analogy of Internet Protocol (IP) packet switching (a network-based approach) has been formulated and is presented in this paper. Using this approach, process orchestration is not required beforehand but can be dynamically specified as the policy-making process is being carried out. This analogy is outlined later in this paper in Section 4. The idea of a network-based approach with a goal-based approach has been put forward by the authors in [15]. This paper further extends the idea and presents our design and an implementation approach in detail which has hitherto not been discussed.

In addition to the above-mentioned requirements, the proposed solution considers the governance principles that form an integral part of policies [13]. Acknowledging smart governance principles in ICT solution also promotes smart government objectives [7,13]. The smart governance principles [7] that have been considered for orchestrating a solution are: (i) citizens' participation in the overall policy process (from the perspective of both "bottom up" and "top down" approaches to policy making) and (ii) facilitating communication among stakeholders (connected silos). These two models have been considered because the impact of public policies is not limited to an organisation (as in many business setups [11]) but has influence on citizens. Thus, policy making requires inputs and suggestions from other concerned authorities. Considering this, good governance in policy-making certainly requires the inclusion of other connected stakeholders (including citizens) [16] and therefore the technology solution must incorporate and support the participation of external and concerned bodies in the policy process. It is to be mentioned here that our proposed system only intends to facilitate communication and involve various stakeholders; our aim is not to improve or propose new principles for smart governance or for smart government. Further details are presented in Section 2.

In summary, public policies play a significant role in smart cities' development [12]. This research considers smart governance objectives and policy domain information for designing a solution to assist government in policy making operations and for tracking the policy process. Policy-making processes are, commonly, lengthy processes therefore fully testing our system in a practical setup imparts delay in fully testing our approach. Therefore, as a proof-of-concept our approach is implemented using a multi-agent technology. A multi-agent system (MAS) is a system that allows the explicit representation of entities, relationships between entities including agents and the environment. A MAS enables the modelling of real-time environments and can therefore easily represent a practical setup [17]. A policy-making environment that involves various stakeholders (who have different roles and responsibilities) and that requires a large communication amongst them can be simulated using multi-agent approach to demonstrate the working environment. Consequently, MASs can represent large human participation and can simulate our network-based approach and are thus considered for implementing the system.

The remainder of this paper is structured as follows: Section 2 discusses the methodological approach, context and background information. Section 3 presents critical reflection on the suitability of workflow technology for tracking policy processes. Section 4 discusses the network and goal-based approach. A detailed design is sketched in Section 5. Implementation details are covered in Section 6. Finally, conclusions are presented in Section 7. The conceptual diagram representing logical flow of the paper is shown in Figure 1.

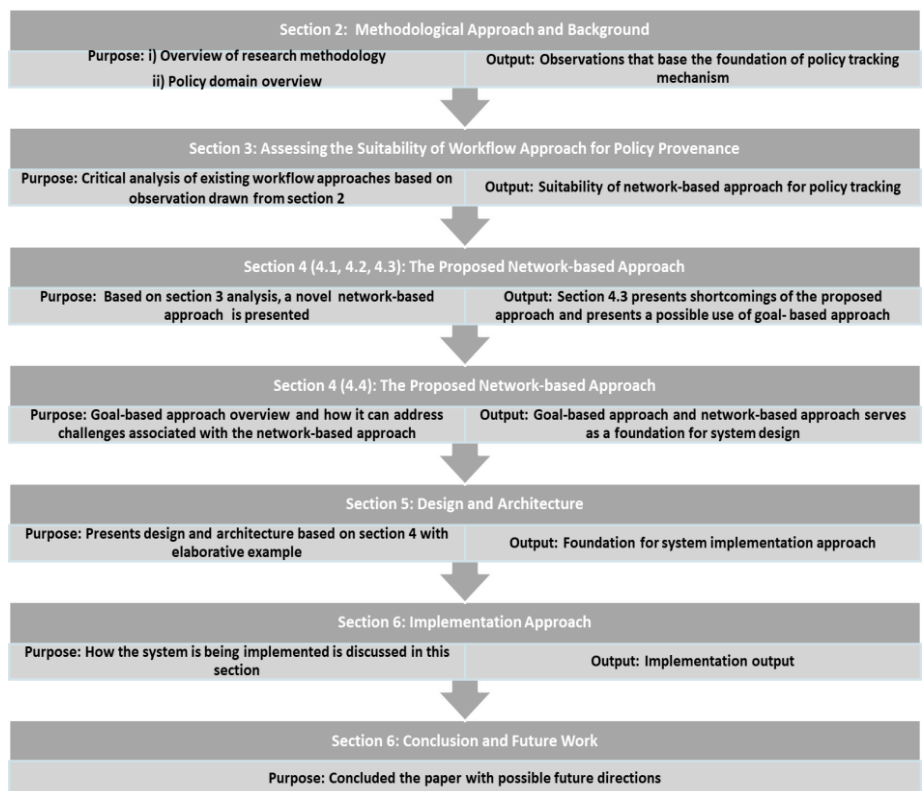

**Figure 1.** Conceptual Diagram of Paper Structure.

## 2. Methodological Approach, Background and Context

*2.1. Methodological Approach*

To design an ICT solution to support and track the policy process, a clear understanding of the domain knowledge relating to policy establishment and how policies are created in a practical scenario is required. For this, the methodological approach we employed was as follows:

(i). We have conducted a detailed investigation of the relevant literature to understand the policy domain and its challenges [11,18–23]; Section 2.2.1 covers the observations in detail. The study was also conducted in the domain of smart cities (details provided in Section 2.2.2) to explore the policy process in smart cities and how the policy process has (or is supposed to) transformed for its suitability in the smart cities context. The investigation uncovered important observations which are apposite to the design of a system that supports provenance management.
(ii). We have elicited input from experts (policy makers) to understand the practical setup of the policy environment and to cross-check the observations that we have drawn from the literature (covered in Section 2.2).
(iii). We have studied existing case studies [18] (presented in the literature) along with other relevant literature [11,19–23] for design and implementation of the system.
(iv). For testing our approach, we are collecting some policy formation examples from the Smarticipate project [24]. It is to be noted here that our system will not be fully deployed in the Smarticipate project; we simply aim at collecting policy examples from the project to verify our approach with the given examples.

*2.2. Background and Context*

To support policy processes and for the policy cycle provenance tracking system, an investigation has been carried from two perspectives as shown in Figure 2. Details of the perspectives are provided in Sections 2.2.1 and 2.2.2 respectively.

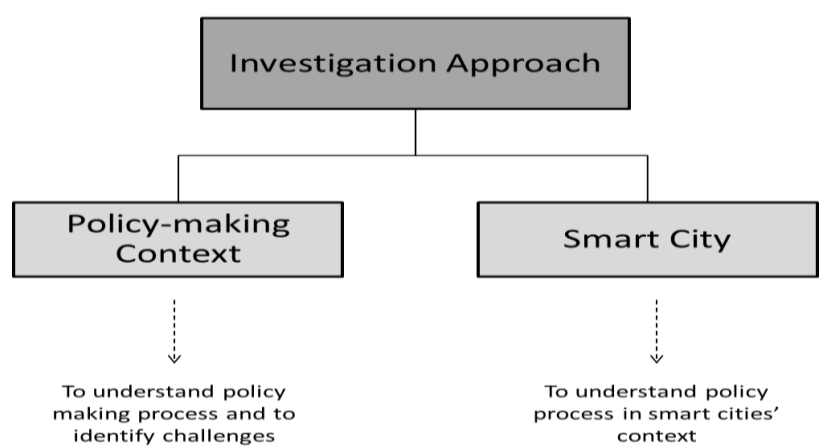

**Figure 2.** Our investigative approach.

2.2.1. Policy Process Domain Knowledge

The design and implementation of a system that supports the policy process and enables policy tracking requires an understanding of the public policy domain. Sajjad [11] covered various existing policy models. Of these models, a generic policy cycle by Macintosh [19] (as shown in Figure 3) seems appropriate to generally explain the policy-making process (refer to [11] for detailed explanation). The generic policy cycle is shown in Figure 3.

This generic cycle demonstrates the life-cycle of a policy-making process; it serves as a framework to devising policy [13]. Our approach uses the generic cycle to frame a solution which is covered in Section 5. Figure 3 demonstrates an over-simplified view of the policy process and only gives an overview of the high-level stages involved in the process but not the internal details of the environment. Thus, for the provision of technology support for the policy-making process, intrinsic details, in addition to the generic model, of policy-making domain are significant. To acquire this knowledge, a further literature survey [11,18–23] has been conducted which uncovered several observations. These observations were cross-checked and validated from the experts in the policy domain. The observations drawn have been categorised into two categories (a) challenges of policy making environment and (b) the characteristics of the policy process.

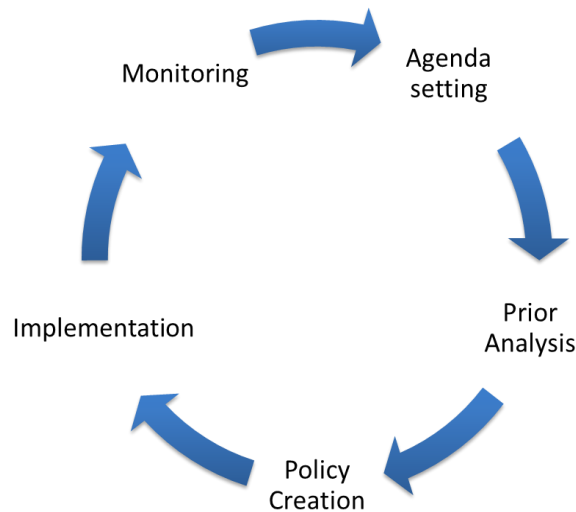

**Figure 3.** The Policy Cycle [11,19].

Domain Challenges of the Process

(i). Although policy-making shares some similar facets with business processes [11] it has been found that, unlike business processes, task specification is not straightforward and can be rather complex in policy-making.
(ii). The process followed for devising policies is largely non-deterministic, unstructured and ad-hoc [22,23] in nature. This means that task specification and task assignment to actors largely depends on the policy and this consequently adds additional challenges in structuring processes. This comes about as a result of policy-making being a political process, with different policy demands and largely involves tacit knowledge of policy-makers to decide the next appropriate action [23].
(iii). The relevant literature [11,18–21,23] states that knowledge intensive aspects are very much prevalent in the process of policy creation. Given this fact, the processes of devising policies are usually guided by human knowledge, experience and decisions. This largely contributes in an ad-hoc manner and leads to complex task specification (covered in points 'i' and 'ii' above).
(iv). In comparison to a business process, the number of stakeholders involved in a policy process is normally significantly large [11,23,25]. These large number of stakeholders require a solution that facilitates cross-organisation communication. This leads to further challenges in process orchestration since actors involved in each policy may not be the same, thereby increasing the complexity of task identification and assignment in each policy. In the case of the specific assignment of roles and responsibilities to actors, the challenge of process orchestration remains because tasks and their sequence for each policy can vary. Furthermore, citizens can participate at any phase which may again increase the complexity associated with the policy process [11].
(v). The policy formulation process is not the same for all policies [23]. Depending on the needs and requirements of policies, the tasks and the relevant stakeholders' participation may fluctuate.
(vi). Policy-making can be unpredictable [23] i.e. it is complex to orchestrate the process beforehand, due to the points 'i,' 'ii,' 'iii,' 'iv' and 'v' outlined above.

(vii). Unlike a business process, a policy-making process may inherently be a lengthy, drawn-out process [11]. Often the execution of any particular task may last from a number of days to a couple of months. One of the factors of this lengthy process is the involvement of a large number of stakeholders. The consultation process with all the stakeholders contribute in lengthy policy-making process.

For designing a solution that addresses the uncovered observations, it is significant to demonstrate any potential overlaps in the identified challenges. Table 1 shows an overall magnitude of challenges contributing in policy process. The table is read from left to right (i.e. challenge given in the left most column is caused by the challenges on the columns to its right).

**Table 1.** Challenges overlap.

| Challenges | Complex Task Specification | Non-Deterministic | Knowledge Intensive | Large no. of Stakeholders (Including Citizens) | Different Policy Processes |
|---|---|---|---|---|---|
| Complex Task Specification | | √ | √ | √ | √ |
| Non-deterministic | | | √ | | |
| Knowledge Intensive | | | | | |
| Large No. of Stakeholders (including citizens) | | | | | |
| Different Policy Processes | | | √ | | |
| Unpredictable | √ | √ | √ | √ | √ |
| Lengthy | | | | √ | |

Table 1 shows that most of the challenges (including complex task specification, their non-deterministic nature, different policy processes and the overhead of prior process orchestration) are due to the knowledge intensive environment of policy formation. The environment demands a solution that acknowledges the human role as a key ingredient of policy-making environment. This is further discussed in Sections 3 and 4. Section 3 focuses on exploring if existing technology solutions can be used to facilitate the process. Based on Section 3, Section 4 covers the approach which has been designed as per the policy process characteristics and its challenges. "Unpredictable" and "Lengthy" do not directly contribute in any challenge therefore these two challenges are not shown in the table i.e. not given in the right most columns of the table.

Characteristics of Policy Process

To devise an ICT solution for the policy-making process, it is vital to understand what the policy process encompasses. Thus, this section covers the process details (characteristics).

Characteristic i: Generally, policy formation encompasses the communication and collaboration within the council, between the council and residents and between the council and partners from other departments; routing tasks; information generation; processing information, storing it and redirecting it to the relevant stakeholders; and information collection by means of various sources such as research, survey and consultation [11].

Characteristic ii: Although the policy-making process is cyclic in nature, as shown in Figure 3, it is not necessary that all phases execute. There may be cases where only a few phases such as agenda setting, or analysis are carried out and the policy-makers decide not to carry the process further.

Characteristic iii: The flow of information between tasks may not be sequential [11,23]; our study and expert suggestion shows that loop(s) back to previously executed tasks or phases may be required as shown in Figure 4.

Our observations demonstrate the operation of the policy environment and highlight inherent challenges. In addition to these, governance plays a significant role in the policy process [13]. For this, smart governance objectives, in addition to the identified challenges, have been considered for designing our approach to facilitate the policy process and to provide a tracking service; this is further elaborated in the following section.

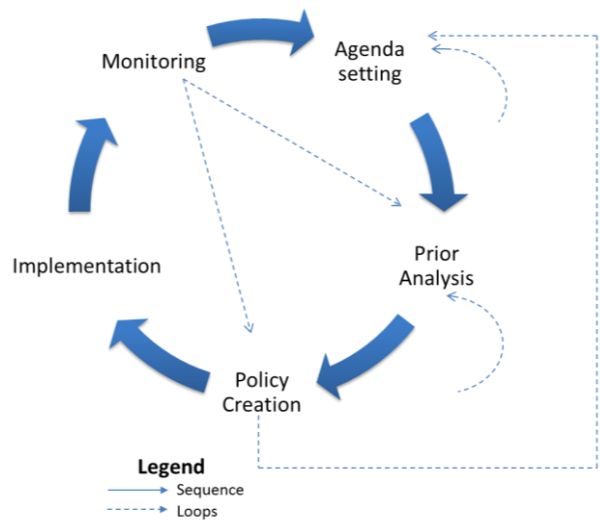

**Figure 4.** Loops in Policy Cycle.

2.2.2. Smart City Concepts

The observations identified in Section 2.2.1 show that the intrinsic nature of the policy process requires the involvement of various stakeholders as it is largely a knowledge intensive environment and human participation, decisions and experience play pivotal roles. Acknowledging the participation and the collaboration of various bodies (government, non-government agencies, citizens) in the decision-making process (which also includes the policy making process) has also been identified as an important aspect for smart governance [7,26]. In this context, we have considered the smart governance dimensions, in addition to the identified challenges in Section 2.2.1, for our system design and implementation. This has been identified as important because the aim of our work is to support the policy process and its knowledge intensive nature demands a solution that must incorporate the participation and the collaboration of various stakeholders; consequently, our system proposes an ICT solution that facilitates the various stakeholders' participation (including citizens) in the policy-making process.

This work employs the objectives of smart governance and ICT to assist local government in their operations of policy creation that also promotes smart government [6]. This is because smart governance is an important building block for smart government [6]. Our detailed investigation revealed various conceptualisations of smart governance, government and cities. Nevertheless, for this research we consider the following definitions:

(a). Among many definitions of **smart governance**, we employ the description by Scholl [6] i.e. that smart governance serves as a foundation for smart government and fosters stakeholders' participation and collaboration.
(b). For **smart government,** we also use the description of Scholl [6] i.e. the use of ICT by government to manage and implement policies and use of smart governance principles for conceptualising smart government.
(c). Our research proposes a solution in the context of **smart cities**. However, as noted earlier at present there exists no standard definition of a smart city and several descriptions fall under this [2]. For our work we focus on smart government, which employs ICT and smart governance objectives, thus providing a solution in the smart cities context.

Figure 5 shows the relationships between smart cities concepts; our focus is on the circles presented in "orange". Figure 6 shows that how these concepts relate to our research. It is important to note here that Figure 5 depicts aspects of smart government and smart governance. All existing explanations of the concepts are not shown here as this figure only intends to highlight the aspects that have been acknowledged in our research.

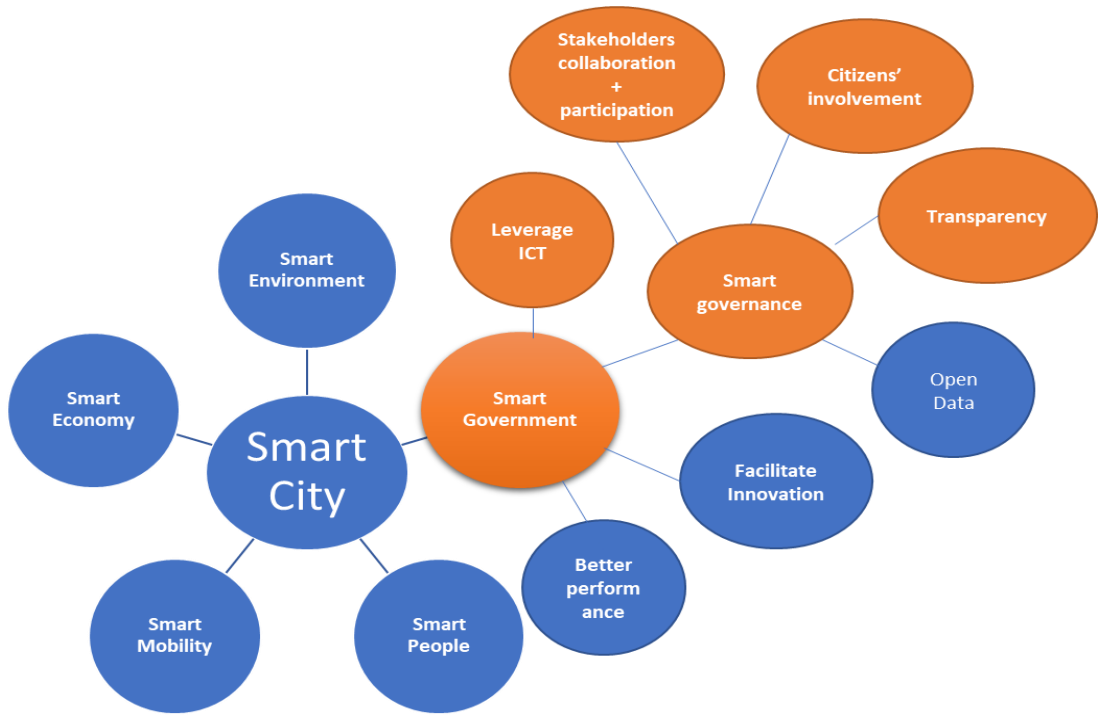

**Figure 5.** The relationship among Smart City, Smart Government and Smart Governance.

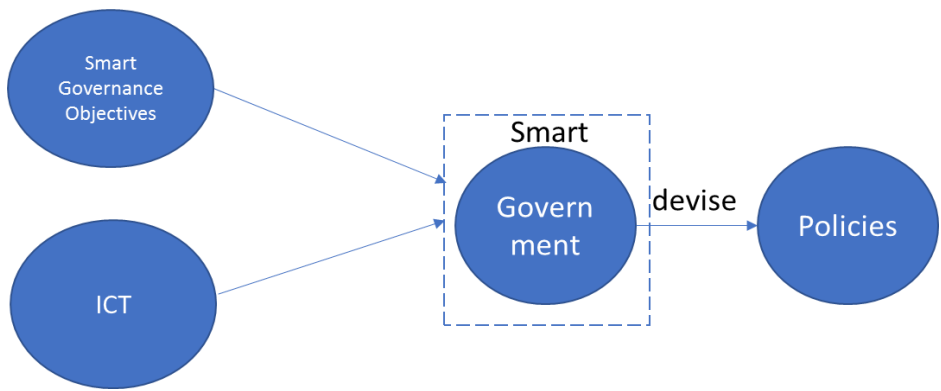

**Figure 6.** The relationship of smart Government and Policy-making.

*2.3. Policy-Making Process in Smart Cities*

To the best of our knowledge, to date there exists no specific policy making process for smart cities. What differentiates smart city policy making from traditional policy making is the governance in the policy-making [27]. Kourtit et al. [28] further demonstrates ICT as a key driver of smart cities policies. Considering this, the generic process (as shown in Figure 3) is applicable for capturing policy processes in smart cities. This generic policy process has thus been taken into account for the design and implementation of our system.

*2.4. Scope of our Research*

(i). Policies are devised at various government levels. Our work is limited to local councils (also called local government) that operate at the city level.
(ii). Our research focuses on smart government and smart governance aspects of smart cities.

To support the policy-making process, it is important to investigate existing technologies that facilitate process and its tracking. Therefore, the next section covers the discussion of existing approaches that have been employed for supporting processes. The discussion helps uncover if the existing solutions are appropriate for policy-making setup.

## 3. Assessing the Suitability of Workflow Approach for Policy Provenance

The observations covered in the previous section set the foundation for choosing an approach that facilitates the process and operation of provenance tracking. For this, we investigated further and found that Sajjad [11] reported on the suitability of workflow technologies for tracking the policy-making process. A significant amount of work has already been carried out in workflow technologies for both scientific and business setups. As the policy-making process is closely allied to the business setup (covered in detail by Sajjad [11]) only business workflows were explored in detail in this research.

*3.1. Potential Workflow Approaches*

From a study of workflow technologies, it has been found that workflows generally fall into categories of structured, semi-structured and unstructured workflows (ad-hoc workflows) [29–31]. However, it has been observed that concepts pertaining to semi-structured and unstructured workflows often overlap and there exists no obvious boundary. For example, Voorhoeve and Van der Aalst [32] present ad-hoc workflows, which are often termed unstructured workflows, in which a process is defined as a procedure that is subject to change at run-time. Some researchers [33] place this definition in the semi-structured approach which collects together flavours of both structured and ad-hoc processes. In order to avoid confusion, we have called these two approaches "flexible workflows"; these workflows enable dynamic changes in business processes. We have done this because both approaches handle the modification (in terms of adding and deleting activity (or any part of a process)) and the construction of processes at run-time. In this paper, we discuss different types of workflows under two categories: (i) structured and (ii) flexible workflows. Our detailed investigation highlighted considerable literature on workflows including structured and flexible workflows.

Structured workflows involve pre-defining processes in detail at design time and then using workflow technology to automate those processes. For this purpose, a number of structured workflow technologies such as Staffware and MQ Series are available that help in defining a process at design time [30]. A substantial amount of work has also been carried out to provide technologies, such as BPMN [34] and BPEL [35], for defining structured business processes. As with structured workflows, a number of flexible approaches for enabling dynamism in workflows have been investigated. For instance, Meng et al. [36] provided an architecture of ad-hoc workflows based on mobile agents and rule-based processing. In their approach, an explicit definition of workflows is not considered but is specified implicitly by business rules. These rules then serve as the foundation to dynamically creating the process. Here the advantage lies with the dynamic modification in business rules which is more appropriate than dynamically changing the well-defined process.

Voorhoeve and Van der Aalst [32] present a definition of a process template and business rules for enabling and controlling flexibility in workflows. The process template serves as a skeleton or template which is modified according to individual cases; the rules are used to enable the changes. Nevertheless, the closer the modifications are to the template, the more accurate the modifications will be. Dustdar [37] employs ad-hoc workflows in a process-aware collaborative system called "Caramba". This project aims at supporting collaboration in virtual teams. Caramba provides a process modeller which users can consider in the case of a known process. Though Caramba supports ad-hoc and collaborative processes, process designers still require an awareness of participants and their intended interactions before process execution. Huth et al. [38] categorised both ad-hoc and semi-structured workflow models in the GroupProcess project. According to them ad-hoc workflows are those workflows that are short-lived, created for urgent cases, exceptional and confidential whereas semi-structured workflows are those that can be predetermined, highly recurrent and can easily enable ad-hoc modification and re-routing. Siebert [39] proposed adaptive workflows that change workflows instances at the run-time. This supports an adaptive approach for handling unstructured processes. The notion of workflow modifications based on previous workflow cases has been proposed by Minor et al. [40]; the authors proposed a case-based approach for workflow adaptation where previous workflow cases are employed to address changes. The changes are dealt with at the run-time and the authors termed such workflows as adaptive or agile workflows.

*3.2. A Critical Analysis of Workflow Approaches*

Both structured and flexible approaches can be used to capture a policy-making process with certain limitations. A well-defined process requires the orchestration of policy-making processes at design time. However, observations covered in Section 2 clearly highlighted the dynamic and ad-hoc nature of the policy process. Consequently, if a structured process is followed then frequent changes are required to facilitate the capture of provenance information. Further exploration uncovered a possible use of ad-hoc workflows in which a process changes and is constructed at run-time. The nature of ad-hoc workflows seemingly addresses the problems of tracking a policy process. Using an ad-hoc approach, processes can be created and added in the executing process at run-time which is very suitable for environments such as policy-making where processes are difficult to outline at the design time. Changing workflows once they are in execution context is however challenging and may require reset of workflow processes [41]. Moreover, discovering exact possible steps, both at run-time and design time, is not always possible [40,42].

Though ad-hoc workflows seem appropriate for policy making but they require knowledge regarding the next required set of tasks. In a business setup, this is not very complex as tasks/activities or processes can often be identified in relation to the modified business rules and regulations. However, this is not achievable for policy-making where the enactment of the next activity/ies depends largely on policy needs and human knowledge and experience (except for the approval from a set-defined body; in this case rules can be specified regarding what approval bodies are required for policies) thus presenting challenge for in-advance process specification. Furthermore, a complicated task specification also makes it challenging to define exactly what execution would be required for the activity. In order to understand this, let us suppose a "review" is required in policy1 to identify possible modifications in existing policies; for policy2, a review may entail collecting *citizens'* comments and analysing them in the light of existing requirements; policy3 may require a *council officer* to visit the site and provide a review report. This clearly shows that the semantics or definition of activities in a policy-environment is very complicated and largely human-centric. Section 2 highlighted that a well-defined process being suitable for all policies was not viable for policy-making; this shows that even if ad-hoc workflows are employed then an overhead of frequent changes will be incurred.

A complicated task specification, variable task definitions, frequent modifications, and challenges in the description of business rules regarding each and every activity of policy setup make existing ad-hoc workflow approaches less attractive. To identify other alternatives for the mentioned challenges, we found that a different approach regarding process formulation, based on human interactions, has been proposed by Sui [43] who employed human actions for the definition of activities. Their approach is inspired from the concepts pinned down by Harrison-Broninski [44]. According to Harrison-Broninski [44], human-driven processes need to consider human interactions and actions for bringing any change into the business process. The authors argued that existing business processes must consider how humans work and not on defining a business process where humans are only considered as a component. The author presented a general overview of current BPM technologies which are mainly suitable for activities executed by machines and lack required human-centred work processes. This idea seems very suitable for the policy domain also where humans are central to the process.

This idea has also been followed by many case management tools [45,46] which are suitable for knowledge work (also called knowledge intensive) environments. These are environments where information plays a central role and processes a secondary role; this approach is suitable where processes are non-repeatable, short-lived, unpredictable, and emergent. Knowledge workers (experts in the domain) decide and orchestrate a process on the spot, depending on the case at hand. The stance of the knowledge workers' environment also holds true for the policy domain because the decision of the next required activity or a process depends on the specific policy case. The available information assists in deciding the next correct activity or required process. Therefore, available case management tools (e.g. [46]) may seem suitable candidates for tracking the policy process. Nonetheless, case management tools focus on the definition of short-lived processes for specific cases but what the policy-making environment suggests is that the next activity largely depends on what information is available. Therefore, defining a process (even if short lived) requires know-how of the next set of activities. However, in the case of the policy environment, where knowledge and information guide the next action, we are presented with difficulties in identifying the precise process and this may leave the possibility of identifying a process that may be subject to further modifications. Moreover, Section 2 showed that processes vary for policies which suggests that if short-live processes are considered then for each policy a process orchestration is required, which is practically not feasible. In addition, a policy may require the definition of multiple short-lived processes due to the non-deterministic nature of the policy process. In order to remove these possible overheads, we need an approach where a process is not a-priori constructed but rather the focus is on the identification of the next atomic activity depending on the knowledge at hand. This will remove the overhead of defining and re-defining a process.

Our investigation revealed that ad-hoc workflows may not fully serve the nature of policy environment. Therefore, we have aligned our approach with Harrison-Broninski [44] ideas by introducing and designing a novel network-based approach for policy tracking in smart cities. The network approach is explained in the following section.

## 4. The Proposed Network-Based Approach

*4.1. The Rationale for a Network-Based Approach for Tracking Policy Cycle*

Policy-making processes have a dynamic nature i.e. the next task selection and execution depends on the information at hand; similarly, the routing of the information to the relevant actor also depends on policy requirements, which vary with policies. The dynamicity associated with the policy-making process therefore demands a dynamic approach to facilitating the non-deterministic policy-making process. Considering this, a network-based approach inspired from IP packet-switching has been adopted for designing our system. Originally, the concept of packet switching was designed to exploit the efficient use of a network and to address concerns regarding dedicated channels for communication. Nevertheless, our aim is to create a dynamic setup but not with the intent of efficient network usage or for employing the same channel for transmitting various communications (as in packet switching) but for supporting an activity-based approach for a dynamic policy-making environment. In relation to this, it is important to identify what similarities and differences a packet switching approach shares with the approach we have devised for tracking; this is covered in the next section.

*4.2. Network-Based Approach*

Section 3 highlighted the limitations of a process-based approach for a policy provenance tracking system. In our approach, the dynamic activity-based approach to tracking policies is achieved by using the IP packet switching technique of computer networks where paths/routes (or channels in network terminology) are not pre-allocated but are apportioned at run-time. Furthermore, messages are carried from one node to another using well-defined message formats which are called packets. The significance of packet switching approach is that packets are created whenever a request or response is initiated; the presence of these packets is otherwise not required in a network. This makes it very suitable for a rapid changing environment of computer network. An interesting point regarding these packets is that a format is predefined but its contents (the data in both header and payload) are decided at run time depending on the required/provided services. Using a well-defined structure allows parties (in this case nodes), involved in communication, to talk using a common vocabulary. The packets hold addresses of involved parties but do not require maintaining the connection during conversation. The packets are sent to their destinations by the routers; routers serve as a controlling component in an ad-hoc network environment. The concepts of packets, nodes, channels, and routers can be mapped to the policy-making process to provide the functionality needed to support the ad-hoc nature of human-centred policy formation.

As with the run-time allocation of channels in a network, a policy-tracking system creates a run-time link between stakeholders; the stakeholders can be analogous to nodes. We define actors as nodes because these actors will be communicating to each other to provide or receive policy details and information. As with the concept of a packet, a token is introduced in our work. The token contains different slots such as IDs, source addresses, destination addresses, request details, the response required from the requested resource or not etc. In computer networks, when a client generates a request the produced information is packaged, labelled, and sent on their way by the IP. This packaged and labelled entity is called a packet. However, in our provenance tracking system IP has no role in place. In this respect, the provenance collection points have been specified in the system which creates tokens and store them in a database. As noted above tokens are the carriers of information. These tokens carry all the details such as who initiated the communication, its destination, the activity details, when it was created, why there is a requirement for the activity etc. In addition to carrying information to destinations, tokens have also been employed to represent the task/activity carried out by an actor (the token templates are further discussed in Section 5.1.3). These tokens present the required *provenance information*. To track provenance, all details pertaining to tokens are captured. Using this approach, the process is reconstructed from the available provenance stored in the database.

Similar to the concept of a router, a "policy controller" (this is a provenance collection point) has been introduced in our system. This controlling component is responsible for managing the policy process; it creates tokens, fills in the relevant details (such as sequence number, time stamp, source address, destination address etc.), routes tokens to the destination which include routing within the council and to stakeholders who are external to the council, receives tokens, stores information, and manages the lengthy policy process. The role of the "policy controller" is further elaborated in Section 5.

The Effectiveness of the Proposed Network Based Approach

The network-based approach demonstrates a dynamic activity-based mechanism for facilitating a policy process, but it is significant to show that how our proposed network-based approach addresses all the identified challenges (given in Section 2). Table 2 presented the effectiveness of the proposed solution by responding to the previously identified challenges. Table 3 shows that how our approach facilitates the policy process characteristics given in Section 2.

**Table 2.** Effectiveness of the network-based approach—handling the identified challenges.

| Challenges (From Section 2.2.1) | Description |
|---|---|
| Complex Task Specification | Network-based approach does not require pre-identification of activities and their sequence. Tokens are created at run-time which contain details of task specified by the actor (Table 1 shows that knowledge-intensive policy-making nature inputs in the challenge of tasks specification). Thus, our approach considers human input in task specification at the run-time. Actors specify the tasks as per policy demands which addresses challenge associated with pre-defining a process. |
| Non-deterministic | Our approach does not consider construction and restructuring of process to guide human action. In our approach, humans guide the process (as non-determinism is due to knowledge-intensive nature as shown in Table 1) which is constructed on the fly from the collected provenance. |
| Knowledge-intensive | No process is defined beforehand but human experience guides the process. Tasks (tokens) are created by policy-makers and routed to the concerned actor. The policy-makers analyses a given piece of information and knowledge at hand to define the next right action. |
| Large no. of stakeholder | In our approach, tokens are carrier of information (this uplifts the challenge of defining a process that spans councils and organisations that are external to the council). Similar to IP packet switching, tokens carry information from one actor to another. While communicating with external authority, the local council keep record of who has been communicated with and from whom a response is |
| | required. No set defined process facilitates inclusion of several diverse stakeholders during the policy formation. Furthermore, this also promotes **smart governance facets** of including diverse stakeholders and citizens during policy-making. |
| Different policy processes | Our proposed approach uplifts this overhead by not considering the process based approach and by taking into account human knowledge as a foundation for policy creation. |
| Unpredictable (difficult to orchestrate process beforehand) | Network-based approach is not process-based thus addressing the challenge of process orchestration. However, policy process reconstruction from provenance data requires efficient algorithms. |
| Lengthy process | Our system maintains state of all policies by tracking activities. These states help reinstate the policy process. |

**Table 3.** Effectiveness of the network-based approach—handling policy process characteristics.

| Policy Process Characteristics (From Section 2.2.1) | Our Proposed Approach Solution |
|---|---|
| Characteristic i | The proposed network approach facilitates communication and collaboration with other stakeholders using a packet like communication. As policy process is largely manual therefore we assume in our approach that someone will have to enter provenance data into the system These details will then automatically be routed (by policy controller) to other stakeholders using the destination address in the packet. All the tokens generated as part of the process are saved in a database which provides an evidence of a process that was executed. All the information that is generated or processed (such as documents) will be stored in a database. |
| Characteristic ii | Based on the human decision, the policy controller decides if policy process continues. |
| Characteristic iii | As process-centric approach is not considered in our proposed solution. Therefore, loops are also not pre-defined. In case loop back is required then as per policy demand, re-execution of previous phase/or certain activities will be carried out. |

*4.3. The Known Shortcomings of the Proposed Network-Based Approach*

Sections 4.1 and 4.2 specified the network-based approach for tracking policies and it can be observed from the discussion that this approach addresses the main challenges (shown in Table 2) but it lacks overall contextual control i.e. process state monitoring in the policy provenance system. This is because if activities are not labelled or specified beforehand, then each actor will specify a name for every activity which will present a real challenge for provenance querying. Using this approach, it will not, a-priori, be known which activity belongs to which phase. However, if humans label activities according to their understanding, it can lead to misunderstandings and usage of diverse and different terms. By using only a network-based approach it will not be known where a policy-making process has reached, what activities have already been executed, and which activities still require execution. Therefore, the system will be very unstructured and uncontrolled. It is important to mention here that provenance ensures transparency in the system but that will be only after it has been collected (*post-factum*). The tokens will be able to provide the provenance of specific activities, but it will not be possible to monitor progress status of the overall policy making process using the specified approach; it will not be able to trace which part of policy process has been executed and which still requires execution. Therefore, to maintain dynamism as well as control (in terms of process state monitoring), a goal-based approach is used alongside the network-based approach, as explained in the next section.

*4.4. Goal-Based Approach*

4.4.1. The Rationale or Using a Goal-Based Approach

Goal-based modelling is not a new concept and has been in use for several years [47,48]. This approach has been employed to identify process and organisational goals. In business, these goals are used to inform the business process. The execution of processes tends to fulfil business and process goals. As discussed in the previous section, having no processes presents other challenges including: (i) No contextual control in the system: by having no well-defined processes, the control in a system cannot be assured. It cannot be known what activities or tasks have been executed and what other activities require execution (ii) It is difficult to specify the start and end and in case of policy cycle, the initiation and termination of phases cannot easily be detailed and (iii) artefacts involved in policy-making process are not known in the absence of well-defined tasks in a process.

These aspects are needed to monitor the status of the policy making process stages and to generate more refined and accurate information for transparency. In addition to these, it also presents challenges relating to provenance which mainly include the efficient querying of provenance. For example, consider provenance tracking using a network-based approach where tokens are the sources of provenance information. In this case, the tokens are carriers of information which captures activity details and takes information from one source to another. Although there is a well-defined structure for tokens (given in the next section) but due to the absence of pre-defined task specification, the representation of tasks by token is not clear. This can be addressed if in a token template, a slot for task specification is provided. This slot can be filled in by actors who create these tokens. This, however, is also not a reasonable approach as actors would be providing their own task definitions; these variable definitions of tasks add no efficiency in provenance querying. Although we are aware that this challenge of variable task specifications may be addressed by using an ontology, however, this is out of the scope of this current research.

The challenges accompanying a network-based approach for policy tracking can be addressed if some structure is added. However, process specification, as discussed previously, for a generic provenance framework is very complex. To cater for this challenge, a goal-based approach with a network-based approach has been proposed. In the goal-based method, goals and sub-goals are identified for each phase of the policy cycle. These goals have been identified from the relevant literature [18–21], for example "problem identification", "validation" etc. More details of these goals are covered in [49]; they provide a high-level view of what is required in each phase of the policy cycle.

4.4.2. Goal-Based Approach for Structuring a Policy-Making Process

In goal-based modelling [47,48], the execution of activities pertaining to a process leads to the fulfilment of these goals. Nonetheless, the identification of the predefined processes that conform to these goals has not been considered; the reason has been already discussed in Section 4.4.1. In this research, the goal-based approach for policy-making has not been employed to identify a process but it has been used to construct a structured process from the collected provenance (using a network-based approach) and for maintaining process states. Table 4 covers details of how goal-based approach covers challenges of a network-based approach.

**Table 4.** The Benefits of a Goal-based approach.

| No. | Challenges of Network-Based Approach | Solution Provided by Goal-Based Approach |
|---|---|---|
| 1 | No process state monitoring in a system | Goals are identified for each phase of policy cycle. These goals guide the policy-makers to carry out the tasks towards goals' fulfilment. |
| 2 | System or process execution is less transparent | Goal-based approach provides an overview of which goals have been satisfied and which goal is currently under execution. |
| 3 | Initiation and termination of policy cycle and its phases | For each goal initiation and termination is specified. |
| 4 | Artefacts and approval bodies involved in a process cannot be stated in an | In policy-making process, approvals at various points are required from the concerned authorities. Therefore, artefacts and approval bodies' information (where required) is associated with each goal of a policy cycle. |
| | ad-hoc environment | |
| 5 | Provenance query | Process for each goal will be created at run-time but activities involved in a process contain ID of goal which associates the process to the goals. For identifying a sequence of tasks, sequence numbers (similar to packets' sequence number) are associated with tokens. Using this approach, provenance can be queried using a goal ID. |

## 5. System Architecture

For assisting the policy-making process and tracking the provenance, a flexible activity-based solution inspired from the IP packet switching has been identified and presented in the previous section. The system architecture, given in Figure 7, has therefore been designed by taking into consideration the network and goal based approach for tracking the policy process. Furthermore, the architecture is based on the Policy Cycle Provenance (PCP) Framework outlined in [49]. As mentioned in Section 2, the generic cycle has been considered as a framework to guiding the policy process and framing our solution. Further it provides a foundation for structuring the activities and provenance. For example, when tokens are created then they include the phase detail which states that which phase is under execution. Similarly, it also assists in querying the provenance; using phase details it can be determined that which activities were carried out during the phase execution.

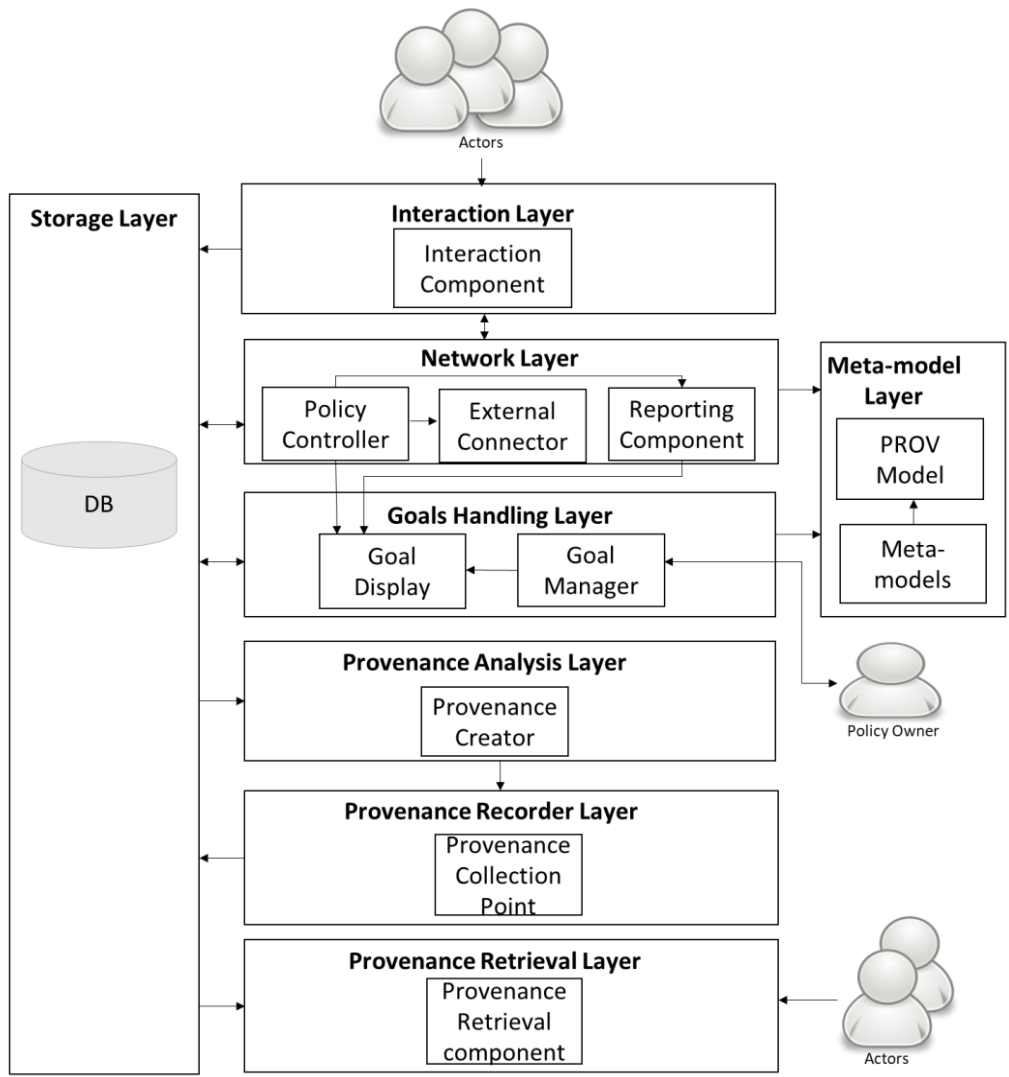

**Figure 7.** Policy Cycle Provenance Architecture.

*5.1. Working of the Architecture*

The architecture has been created by considering the PCP framework presented in [49]. From Figure 7, there are eight main layers in the architecture:

(i) the Goal Handling Layer implements the goal-based approach
(ii) the Network layer implements the network approach
(iii) the Interaction layer provides the interface with the actors
(iv) the meta-model layer provides all details regarding the process, data, structure, provenance details. This layer uses the W3C PROV model [50] to specify the structure of the provenance information
(v) Provenance analysis layer creates the provenance using tokens and as per meta-model layer
(vi) Provenance recorder layer records the provenance
(vii) Provenance Retrieval Layer is responsible for retrieval and display of provenance and
(viii) Storage Layer that stores data and provenance (which we call as data store and provenance store). This layer is used by all layers either for storing or for retrieval of data.

5.1.1. Goal Handling Layer

The Goal Handling Layer comprises two components namely: The **Goal Manager** and the **Goal Display Component**. Before going into the detail of the components, it is important to mention here that the human-centric nature of the policy-making process demands human intervention in the operation of the process. For this purpose, a policy owner is assigned to each policy; this idea has been determined from [13] where each policy case was designated a policy owner. In our system, the policy owner will be a person/department who will be assigned responsibility for managing policies and identifying the metrics of goals for each policy. The goals and metrics are input to the system by using the **Goal Manager** component.

Metrics are specified for each goal for the purpose of monitoring goal execution. The formulation of goal metrics will largely be based upon policy needs and organisational structure. For example (in the case of policy needs), when a new policy is initiated then the policy owner will input what each goal is required to deliver such as a specific parliamentary bill, any particular document etc. The owner will also consider the organizational structure to specify constraints pertaining to the authorising body. For instance, a policy owner can specify which policy-makers/stakeholders will be involved for approval of the policy. Due to the subjectivity and dynamism involved in the system, the goals and metrics are continuously monitored by the **Goal Manager Component**. All the goals and their status are displayed by the **Goal Display Component**. These goals provide the status of the policy which also direct policy makers. For each goal, the specified metrics is stored in the **database** for maintaining record for each policy. The potential use of this component is further explained in Section 5.1.3.

5.1.2. Interaction Layer

This layer focuses on gathering the input from the actors. This layer not only accepts inputs to initiate the process, but it also takes inputs from the policy-makers and other stakeholders while the policy execution is in the process (this is further elaborated in Section 5.2). The interaction component provides an input to the network layer.

5.1.3. Network Layer

The **Interaction Component** in the **interaction layer** provides an input to the **Policy Controller** component in the network layer, as shown in Figure 7. As mentioned in the previous section, the idea of a policy controller is similar to the router which monitors and maintains controls in an ad-hoc environment. For the policy-making environment, the policy controller is responsible for the creation of tokens; routing it to the relevant stakeholders within local council; and in the case of stakeholders external to the council it forwards the details to the **External Connector**; managing sequence numbers, timestamps and recording of the time duration required for any activity execution. It reads the **Goal Display** to include goal details during token creation. Once any goal or any of its metric gets fulfilled then a notification is sent to the **Reporting Component** which updates the **Goal Display**. In case of a new policy, the **Policy Controller** takes this input and informs the policy owner who will then create the metrics for that particular policy. The owner can also modify any goals and goals metrics, if and when required. The **Policy Controller** gets an update when the **Goal Display** component displays the goal and metrics information. For provenance creation, the **Policy Controller** creates tokens and save them in the database. The **Provenance Creator** in **provenance analysis** layer will take the tokens and will create the provenance according to the provenance specification in the Meta-Model Layer. The tokens represent activities/tasks performed by the actor and they are also carriers of information i.e. they carry the data to another actor. Tokens, as mentioned earlier, carry information from one node to another. In addition to this we are also interested in capturing the activities' provenance that are executed by policy-makers. For example, if a policy-maker performs three consecutive activities (we call this a "local-token" in order to differentiate tokens and local-tokens) then this provenance also requires capturing. The local-token and token templates are shown in Figures 8 and 9.

| Policy ID | Phase ID | Goal ID | Seq num | Task | Task Input | Nature of Task Input | Task Output | Nature of Task Output | Task Description | Who Carried out? | Nature of Stakeholders | Data Details | Source of Evidence | Data Reference | Time stamp |
|---|---|---|---|---|---|---|---|---|---|---|---|---|---|---|---|

**Figure 8.** Local Token Template.

| Policy ID | Phase ID | Goal ID | Seq num | Previous Task | Previous Task Output | Nature of Task Output | Task Description | Any required action? | Communication pattern | Nature of Stakeholders | Data Details | Source of Evidence | Communication with? | Data Reference | Time stamp | Source Address | Destination Address |
|---|---|---|---|---|---|---|---|---|---|---|---|---|---|---|---|---|---|

**Figure 9.** Token Structure.

Figures 8 and 9 demonstrate tokens details. Figure 8 shows that local token carries details including task description, input, output, and the nature of task input/output. "Nature of task input/output" depicts the specification of the input and output such as if the task input and output are data, documents, decisions, or any required actions. "Task description" provides details of the task and represents "Why" aspect of provenance. The field "who carried out" represents "Who" of provenance. For tracking record of participating stakeholders, it is significant to add whether those stakeholders are external or internal to the council. This information is required because organisation structures can change anytime. It is possible that a new department may be formed and old departments may no longer exist. Thus, having only the department name may create confusion if a provenance is analysed following a long-time span. For this, Figures 8 and 9 hold the nature of the stakeholder/department field in the token structure. "Source of evidence" represents "how" and "where" of provenance.

Figure 9 shows the structure of token. It carries details of previous task. "Any required action" in Figure 9 shows any required action by the receiver. For example, 'sender A' requires approval from 'receiver B'; in this case the field "any required action" carries detail that approval is required from 'receiver B'. "Communication pattern" depicts communication details (these are discussed in detail in section 5.1.4). "Communication with" represents details of an actor with whom the communication is carried out. "Source and Destination addresses" represent source and destination details of senders and receivers.

Figure 7 shows that the **Policy Controller** communicates with the **External Connector** when communication is required with stakeholders external to the local council. This component holds the addresses of all the external partners and manages the communication with them. As identified in Section 2, the communication with various external stakeholders is important during policy creation which suggests that the network is required to span across various external partners. For design simplicity, each external partner has been considered as a separate network having its own departments, controller, and external connector. The communication among network is facilitated by **External Connectors** that have been specified separately for each network. The **External Connectors** carries token to another network which is handed to the **External Connector** of the destination network. The **External Connector** receive tokens and forward it to the concerned department.

5.1.4. Other Layers in the Architecture

The Provenance Analysis Layer according to the meta-model layer creates the provenance. The "Provenance creator" component in Figure 7 represents the "provenance collector" and the "provenance decision maker" of the PCP framework [19]. The Meta-Model Layer is responsible for specifying the structure, process, and data flow information. It specifies the constraints among different entities. Furthermore, it uses the PROV model [50] for provenance specification. Tokens carry details such as Why, How, Who, When of provenance (as discussed in the previous section). The tokens represent the provenance details which are used to construct provenance using a PROV model. The Provenance Recorder Layer is responsible for recording the provenance in the provenance stores and in the Provenance Retrieval Layer the **Provenance Retrieval Component** retrieves the provenance information from the provenance store in storage layer and displays it. In ongoing work, we are developing a dashboard that will display the queried provenance information.

The conceptual model of the devised approach is given in Figure 10. From Figure 10, it can be seen that tokens hold both process and data details. The process holds information such as the event, control details and different communication patterns which include sequential, parallel, synchronous, and asynchronous. These patterns depict the possibilities of control flow in the system. For example, "Sequential" shows the control is transferred from one actor to another in a sequential manner. "Parallel" on the other hand shows that a control can be passed to multiple actors in parallel for any task execution. "Synchronous" depicts that the next activity cannot be executed until the response has been received by an actor from some other (internal) actor. On the other hand, "Asynchronous" depicts that the next activity can be executed by an actor while awaiting a response from another actor. Furthermore, the combination of communication patterns will also exist such as synchronous-parallel, asynchronous-parallel etc. The examples of sequential, parallel, synchronous, and asynchronous communications patterns are shown in Figure 11.

Figure 10 also shows that tokens will contain both process and data details. This is required because provenance will not only represent the execution details but will also specify the data that is used or created in the process. Processes will include information such as activities' sequence, communication patterns and so on. Data includes details such as user details, input of activity, output of activity, documents etc.

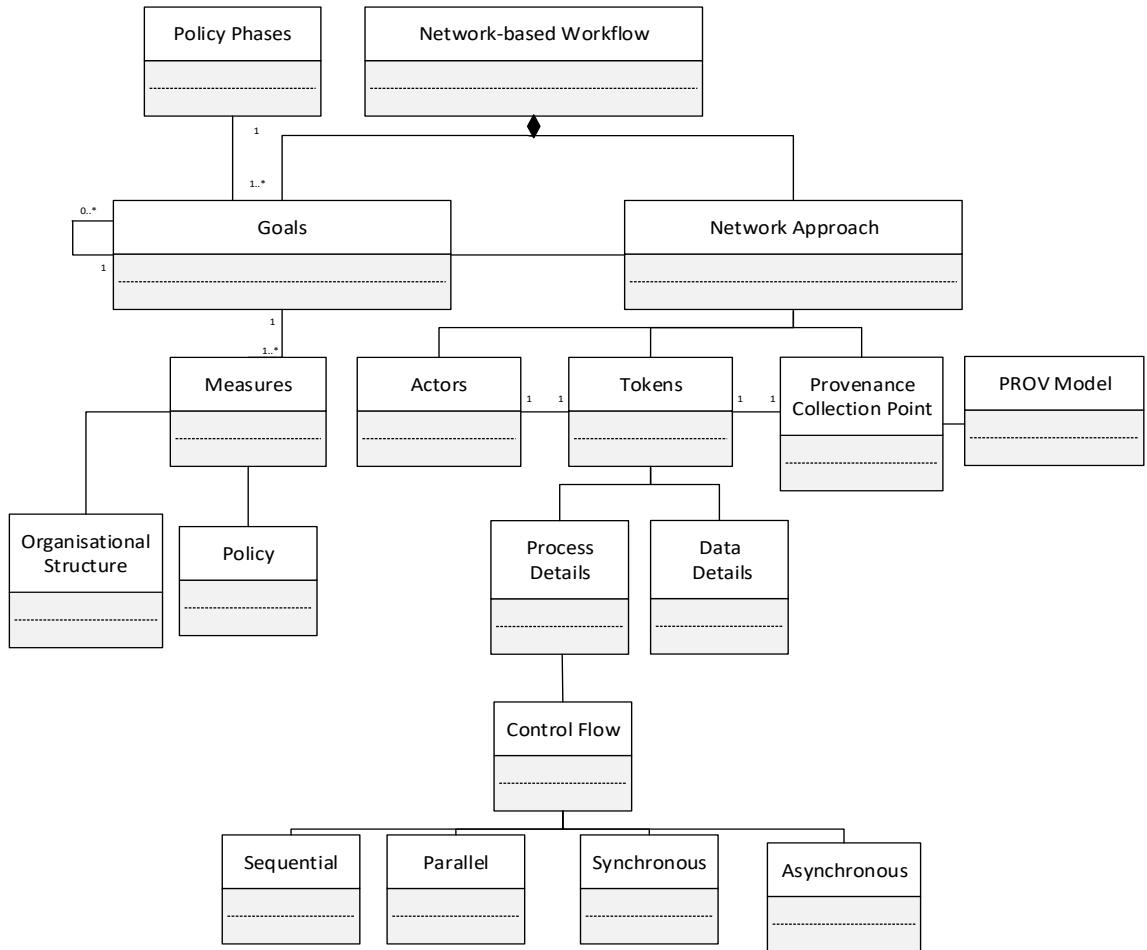

**Figure 10.** Conceptual model of devised approach.

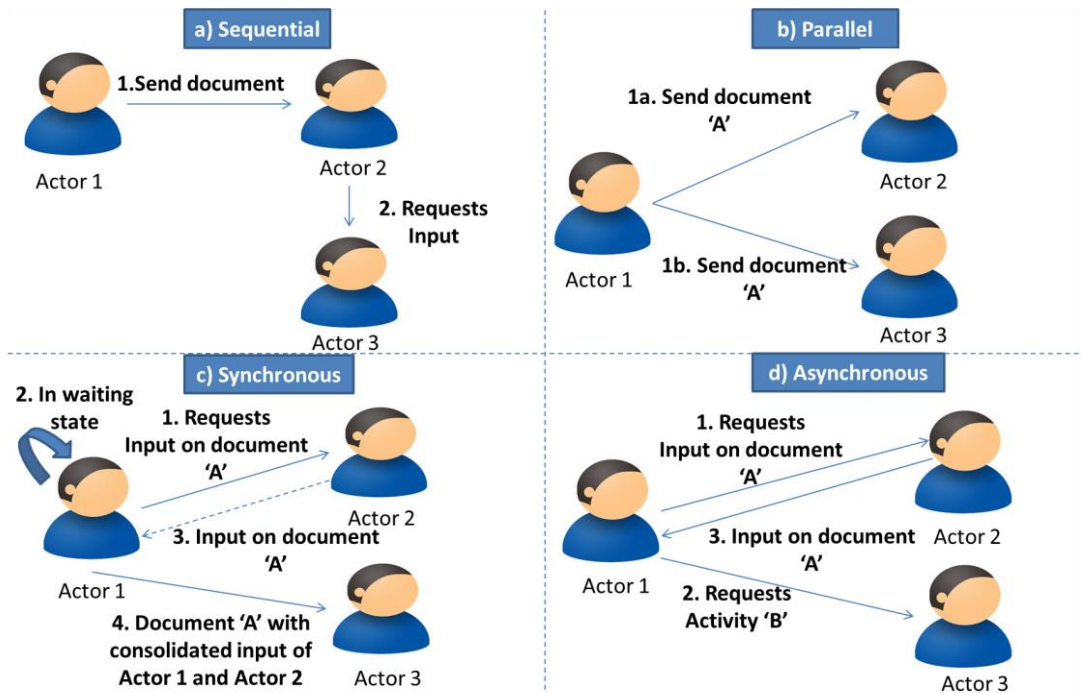

**Figure 11.** Communication Patterns.

*5.2. Example to Demonstrate Functioning*

In order to explain the functioning of our architecture, we have considered, as an example, one of the case studies covered by Tsohou et al. [18]. The case study that we have selected is the "Neighbourhood Safety Planning Process" [18]. In the given case study, the local community generates new ideas; these details are input to the system by using the **Interaction Component** in the **Interaction Layer**. The interaction component is coupled with the **Data Store** in **Storage Layer**. The data (in this example, we assume data are documents) are stored in the **Data Store** and the data notification is sent to the **Policy Controller** component in the **Network Layer**. The **Policy Controller** components informs the policy owner. The policy owner checks the requirements of the required policy and, based on neighbourhood policy requirements, the owner identifies metrics for the goals. Now suppose the policy-owner identifies that for a "problem identification" goal of the "agenda setting" phase, two metrics are required (a) a "problem analysis document" and (b) verification of that document by the "Business control unit (from the case study)". In this case the policy owner does not specify the process that will be followed for the document creation. This is because network-based approach does not require beforehand process orchestration. The identified metrics are provided to the system via the **Goal Controller** component. The "Problem identification" goal is then displayed, on **Goal Display** component, with its two-associated metrics (the problem analysis document and the verification by the business control unit). The **Policy Controller** reads the **Goal Display**, create a local-token, and store in database.

The Policy Controller then notifies the relevant policy maker; in this example it is the "safer neighbourhood team". The communication between the safer neighbourhood team and the business control unit is carried out using a token (in an analogous manner with network packets as detailed earlier) for the transmission of details to the safer neighbourhood team. From here, we can see that the pre-defined connection amongst actors and activities is not required (as network-based approach has been designed in such a manner). On receiving the notification, the safer neighbourhood team starts its activity. Following the case study, the team considers existing policy documents (activity inputs) to identify the list of proposed amendments (activity outputs). Then it reads the **Goal Display** and prepare the local-token which is then stored in a database.

On completion of their task the safer neighbourhood team sends the details of the previously executed activity including a list of proposed amendments to the "business control unit" using a token. The business control unit, on receiving the details, prepares the change management document; the details of which are stored in the database.

According to the case study, the "information gathering" from the "neighbourhood watch team" is required. The request from the business control unit to the neighbourhood watch team is facilitated by tokens (similar to previous steps). The tokens will then form the basis for the capture of provenance information. The neighbourhood watch team then provides the input using an **Interaction Component**. The details will be stored in the Data Store and the provenance of the activity will be saved in the Provenance Store. On receiving the input, the business control unit prepares the "problem analysis report", thereby fulfilling the first goal metric. The analysis details are forwarded to the safer neighbourhood team who then prepare the first draft. The policy draft is subsequently shared with the business control unit for verification. This fulfils the second metric thereby fulfilling that goal. The **Reporting Component** then sends the notification to the policy-owner, via the **Goal Display**, that makes the decision regarding the next goals and measures.

The working example of the network-based approach coupled with goal-based approach is shown in Figure 12.

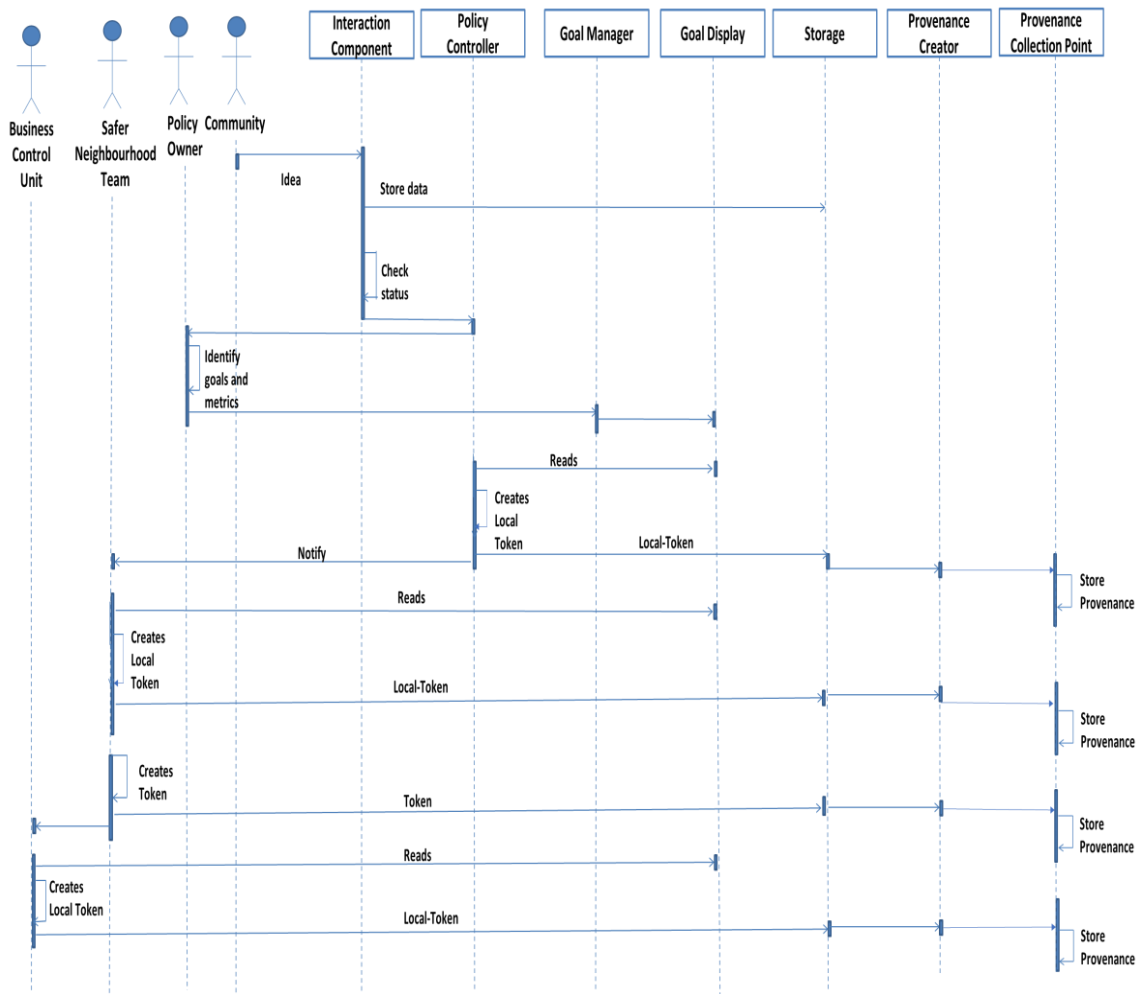

**Figure 12.** Working Example.

## 6. Implementation Approach

For our prototype implementation, we have used a Multi-Agent System (MAS) for simulating the real-time environment of the policy formation process. We are developing the system using JADE [51]. A MAS has been identified as being suitable to facilitate the dynamic and distributed setup. Furthermore, agents can represent users and can work on behalf of them [52]. The policy-making process is dynamic in nature due to processes varying with each policy. In our simulation, agents represent humans involved in formulating policies. These agents interact with each other to devise policies. However, we are not using MAS to actually create a system that assists policy-makers in devising policies, rather we are using MAS to simulate the policy environment to test our approach.

As stated previously, the policy-making process involves various stakeholders including the local council, national government, regional government, various partners (e.g. Health sector), and citizens. For different stakeholders, we have created various networks. Each network represents a different decision-making body. For example, we have created a network that represents the working of a local council, a network of citizens, and a network for each partner involved in the process. The networks demonstrate how agents communicate within the network and with external networks. Section 2 detailed the reason for considering policy-making at local council level. Taking this into account, our simulation represents the details of a local council i.e. its departments, the routing of tasks among internal departments, its connection with external stakeholders (regional, national governments and other partners), and citizens. Due to the distributed nature of the policy formation process, data governance is a challenge. In our simulation, we suppose that data and provenance are stored by the local council. The simulation approach is demonstrated in Figure 13.

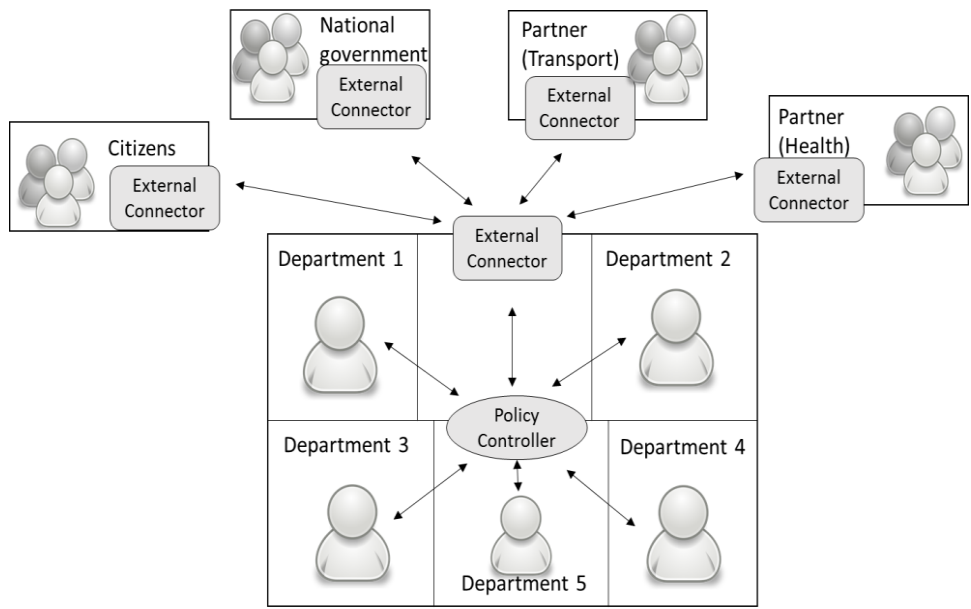

**Figure 13.** Simulation Approach.

Figure 13 depicts how an agent interacts with the system, the control being provided to the policy controller which is dedicated for each network. The policy controller is responsible for creating tokens in the system, storing the tokens, routing the tokens to the destination, keeping a record of all the executing policies, maintaining a record of all internal departments, and the addresses of the external connectors (external connectors are used to connect different networks, the communication between networks being managed by connectors). These connectors maintain details of all the associated networks and manage the communication among the networks. The communication with other stakeholders and citizens is carried out with the help of the external connector(s). These connectors maintain a record of all stakeholders. When a communication is required with external stakeholders then the policy controller passes the details to the external connector; these networks communicate with each other using external connectors. The controller and connectors help serve the provenance collection points and these controllers and connectors also track if goals have been satisfied for maintain the process state.

Our network-based approach supports dynamic setup of policy-making. For testing this, we gave various inputs to the system. The system outputs are shown in Figures 14 and 15. Figure 14 demonstrates the communication of various agents. For example, it can be seen from the figure how the policy controller controls the communication between the "safer neighbourhood team" and the "business control unit". The "safer neighbourhood team" interacts with the system and its input is provided to the "policy controller" which creates the token and forwards it to the "business control unit". We gave different inputs to the system to check how the system responds in a dynamic policy-making environment. Figure 15 shows that this policy process has a different stakeholder (change control board) who is involved in the policy process. In this policy, the "safer neighbourhood team" contacts the "business control unit" which interacts with the "change control board" through the "policy controller". Similarly, other inputs were provided to the system to check how it track different policies. The results of the above simple simulation are encouraging and demonstrate that the network-based approach is suitable for a dynamic policy-making. Our future work will be on testing the above approach with complex scenarios and to identify strengths and limitations of our approach.

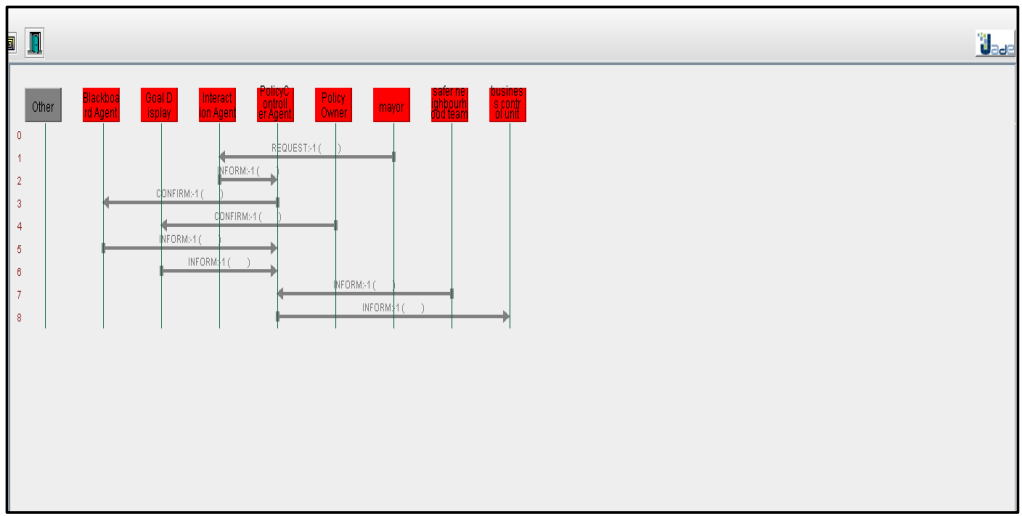

**Figure 14.** System Output-Communication among Stakeholders.

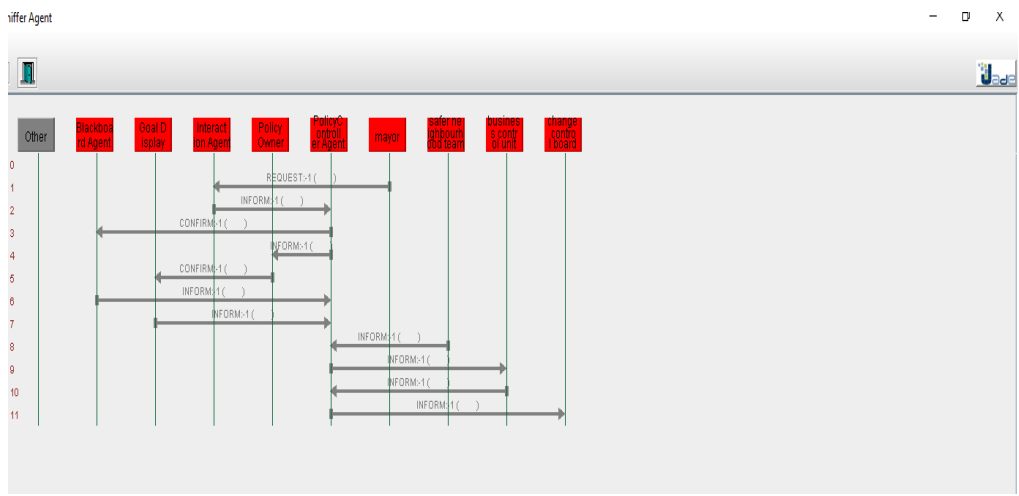

**Figure 15.** System Output- Different Process.

## 7. Conclusions and Future Research Directions

The process data that is used for and in the creation of policies, if tracked, can provide valuable (provenance) information associated with decision choices. In this respect, the research presented in this paper aims at collecting the provenance information of the policy process in smart cities. However, policy making is a complex process and varies in different settings and hence necessitates an ICT-enabled process agnostic approach for tracking policy creation. To this end a detailed investigation has been carried out and a number of observations have been drawn that assisted in establishing an architecture and the working of a provenance system for policy process tracking. The architecture of the provenance capturing system has been designed by introducing a novel network and a goal-based approach, as detailed in this paper. These approaches were identified after investigating available literature on workflow technologies. The observations regarding the policy-making process highlighted that current workflow approaches are not suitable due to the large amount of manual activities, lengthy task execution, and the non-determinism in terms of tasks details and process. The designed network-based approach employs concepts drawn from the network domain to facilitate the policy-making process.

This research will benefit policy-makers in terms of providing a system that can work in an ad-hoc and political environment of the policy-making in cities. This approach provides flexibility in terms of collecting data regarding diverse policies. Using this system, councils can define their own processes without need to consider the challenges associated with workflow formulation. Note that the approach taken in this paper, whilst using 'Smart Cities' as the test ground for policy making is generic in form and design. The smart governance objectives including active citizens' participation and multiple stakeholders' involvement have been considered in the system implementation. Our approach does not intend to propose new governance concepts or improve the governance for smart cities; we focus in using the existing smart governance models in system design and implementation. This research does not aim at developing a system for a real-time setup due to time constraints. Therefore, the design of the network-based and goal-based approaches is tested by developing a light-weight proof of concept in Multi-Agent Systems. Different process inputs were provided to the system to check if the proposed approach can handle various policy processes. The testing demonstrates the suitability of the network-based approach for a dynamic policy environment. We are continuing to investigate other means to test and validate our approach.

Software agents can learn over time to decide the next action, but this is out of scope of the current research. Future work will consider testing this approach by considering practical examples. In this respect, new data is being collected from the H2020 Smarticipate project that will be used to verify and validate the provenance approach. With network-based approach, ontology can be used to define terms that will assist in process specification.

**Author Contributions:** BJ conducted the research and wrote the paper. ZK and RM introduced the idea of provenance for policy making cycle and provided guidance in developing research idea and in formulating the paper.

**Conflicts of Interest: The authors declare no conflict of interest.**